# Machine Learning Method Used to Discrete and Predictive Treatment of Cancer

**SeyedMehdi Abtahi[1], Mojtaba Sharifi[2,1,*]**

[1] *Department of Mechanical and industrial Engineering, University of Illinois at Chicago, Chicago, Illinois, 60607, USA*

[2] *Department of Electrical and Computer Engineering, University of Alberta, Edmonton, Alberta, T6G 1H9 Canada*

# Machine Learning Method Used to Discrete and Predictive Treatment of Cancer

**Abstract**

Cancer is one of the most common diseases worldwide, posing a serious threat to human health and leading to the deaths of a large number of people. It was observed during the drug administration in chemotherapy that immune cells, cancer cells and normal cells are killed or at least seriously injured and also in order to keep dosage of the drug at specific level in body, drug should be delivered in specific time and dosage. Therefore, to address these problems, a decision-making process is needed to identify the most appropriate treatment for cancer cases which causes killing of cancer cells by considering the number of healthy cells that would be killed. Despite the latest technological developments, the current methods need to be improved to suggest the most optimized a dose of the drug for tumor cells discretely. In this paper we proposed mathematical models for tumor, safe and infected cells and also fat in patient body then by these coupled equations we are able to train our model via using Machine Learning method. It is expected that our proposed ANFIS model be able to suggest the specialists the most optimum dose of the drug, which considers all key factors including cancer cells, immune and health cells. The results of the simulations exhibit the high accuracy of the proposed intelligent controller during the treatment in predicting the behavior of all key factors and minimize the usage dose of the drug with regard this significant point that the proposed controller gives discrete data for treatment which can fill the gap between engineering and medical science. Conclusion is that, using of Machine Learning method could help us to find the shortest period of treatment for each patient and also monitoring the other important factors during the treatment.

## 1. Introduction

Despite a huge research carried out in this domain, the prediction and treatment of cancer is very difficult, and in some cases impossible. In fact, a wide variety of factors play a key role in predicting and treating cancer, and more importantly, they have different effects on each other. Colorectal cancer incidence rates in men and women are 37.8% and 27.8%, respectively. It should be noted that cancer mortality in men and women is 12.4% and 8.8%, respectively [1].

This current complexity and above all the static make it necessary to understand changes in the dynamics of all-important and effective cancer states. To eliminate, reduce, or at least stabilize the number of tumor cells, the best controllers should be developed and used [2].

Over the last two decades, various control methods have been used to target various cancers. Florian et al. [3] studied a nonlinear model predictive control algorithm for breast cancer treatment. For example, an adaptive robust control strategy is developed for the manipulation of drug usage and consequently the tumor volume in cancer chemotherapy by Moradi et al. [4]. Yazdanpanah Goharrizi et al. [5] studied a self-tuning adaptive controller for 3-D image-guided ultrasound cancer therapy.

Optimal controllers have been widely used to simultaneously control tumor and the amount of drug usage [6]. An overview of the way optimal control theory interacts with cancer chemotherapy was presented by Swan [7]. a schematic demonstration was brought forward by Moradi et al. [8]. In their study, optimal robust controller for drug delivery in cancer chemotherapy was sought and the performance of three control approaches was compared. In the other work, Batmani and Khaloozadeh [9] worked on optimal chemotherapy for cancer treatment, they used SDRE in order to eradicate tumor. The main drawback of the optimal control devices is their inability to predict the required drug when the uncertainties are high [10].

One way to overcome this disadvantage is through integrated controls, including optimal and fuzzy methods. Fuzzy logic has been used to help researchers find effective treatments for cancer. Al-Daoud [11] studied cancer diagnosis using a modified fuzzy network. Evaluation of breast cancer risk using fuzzy logic has been reported by Balanica et al. [12]. In another similar research El Hamidi et al. [13] worked on an evolutionary Neuro-Fuzzy approach toward diagnosis of breast cancer.

In this paper, an ANFIS (adapted neuro-fuzzy inference system) [14] controller is presented to treat the cancer. This controller can work with unstructured uncertainty. The other great ability of this controller is that it allows us to provide treatment schedule with discrete data, which represents a significant improvement in biomechanics, as this ability can connect engineers directly to specialists. The other advantage of this important ability follows the role of all states in the mathematical models. In addition, this model offers specialists to discreetly check the other important conditions at any point during a treatment phase. Most importantly, however, that the previously proposed controller needs feedback from all states, and here are the biggest differences between specialists and engineers in dealing with this problem, but in our system to find out what dose of the drug you just use the amount of the tumor for the first time, the proposed ANFIS system provides all the important data needed for the discrete treatment of cancer.

The rest of this paper is organized as follows. Sec. 2 presents the stablished mathematical model of cancer. The structure of PID optimal controller is described in Sec. 3. The proposed

ANFIS structure is given completely in Sec. 4. The structure of the ANFIS controller and the ANFIS observer is described in the sec. 4.2. In Sec. 5, the simulation results are demonstrated and compared with those obtained from a previous method. The completion of this work is expressed in Sec. 6.

## 2. Mathematical Model of HBV

There are many approaches in order to find the mathematical model for tumor cell treatment. Shabel and et al. [15] did the first step in this way and consider cancer cell killing rate of a drug is proportional to the tumor population. Norton and Simon [16] could not explain clinical results in some cases, and they found out the cell killing rate is dependent on tumor growth rate. In 2012, d'Onforio and et al. [17] showed that the Norton and Simon's model can produce appropriate results in comparison with experimental observations.

In this paper, the basic model presented by [18, 19] is considered, this model contains five states, immune cells (I), tumor cells (T), normal cells (N), obesity (F) and chemotherapy effect (M). The nonlinear system of differential equation is presented here

$$\frac{dI}{dt} = s + (\frac{rIT}{a+T}) - c_1 IT - d_1 I - a_1(1 - \exp(-M))I \quad 1$$

$$\frac{dT}{dt} = r_1 T(1 - b_1 T) - c_2 IT - c_3 NT + c_5 TF - a_2(1 - \exp(-M))T \quad 2$$

$$\frac{dN}{dt} = r_2 N(1 - b_2 N) - c_4 TN - a_3(1 - \exp(-M))N \quad 3$$

$$\frac{dF}{dt} = r_3 F(1 - b_3 F) - a_4(1 - \exp(-M))F \quad 4$$

$$\frac{dM}{dt} = u - d_2 M \quad 5$$

All parameters, their symbols and default values used in our purposed model are presented in table (2)

**Table 1** Parameters value

| Parameters | Value |
|---|---|
| $a_1$ | 0.2 |
| $a_2$ | 1 |
| $a_3$ | 0.1 |
| $a_4$ | 0.3 |
| $b_1$ | 1 |
| $b_2$ | 1 |
| $b_3$ | 1 |
| $c_1$ | 1 |
| $c_2$ | 0.5 |
| $c_3$ | 1 |
| $c_4$ | 1 |
| $c_5$ | 1 |
| $d_1$ | 0.2 |
| $d_2$ | 1 |
| $r_1$ | 1.5 |
| $r_2$ | 1 |
| $r_3$ | 1 |
| $s$ | 0.33 |
| $r$ | 0.01 |
| $a$ | 0.3 |

## 3. Optimal PID Controller

The amount of necessitate drug by using PID controller can be computed as:

$$u = k_p T + K_I \int T \quad 6$$

Optimizing the control gains ($k_p, k_I$) of equation (6) so as to minimize our cost function:

$$J = \int |u| dt \quad 7$$

An appropriate cost function is defined such that the amount of required drug is minimized while the tumor volume is reduced.

In order to guarantees the smooth behavior of drug some constraints are considered.

## 4. Adaptive network-based fuzzy inference system

### 4.1 Fuzzy

The combination of the neuronal network and fuzzy with allowing for neuro-fuzzy control. As the ANFIS design begins, the least squares gradient decay is used for the training. Sample data from our optimal PID controller is used to train the ANFIS controller. Each learning level is divided into two parts. In the forward stage, optimal coefficients are provided using the inputs and outputs of each layer. In the reverse phase, the parameters of the ANFIS system are then updated. The goal is determined based on the ideal response of the system under step input. ANFIS only supports Sugeno [20] systems.

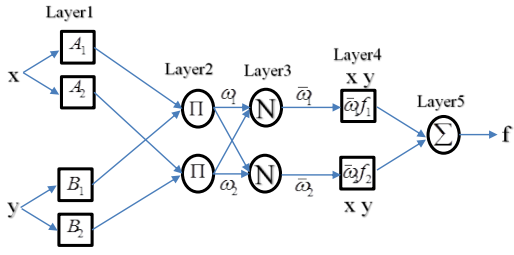

**Fig 1.** ANFIS Structure

Layer 1: degree of membership of the input is computed
Layer 2: the incoming signals are multiplied.
Layer 3: output of layer 2 is normalized.
Layer 4: output function and output of layer 3 are multiplied.
Layers 5: summation of all outputs in layer 4.

**4.2.1 ANFIS controller**
In this paper, the aim of training the ANFIS controller is to discreetly predict the dose of the drug whenever the specialists deem necessary, by measuring the amount or volume of the tumor cells alone on the first day of treatment. To fulfill this purpose, an optimal PID regulator has been designed by regular input variables, including immune cells, tumor cells, normal cells, obesity factor and chemotherapy. As a result, this system will deliver a dose of drugs appropriately with tumor cells during the treatment period.

Our practical ANFIS controller is able to show the specialist the amount or volume of tumor cells and the correct dose of the drug during treatment at each stage of the treatment they need. More importantly, it can help them to predict the behavior of the tumor cells under different conditions (with the different dose of the drug), this capability makes specialists able to treat each patient in terms of his or her drug reactions.

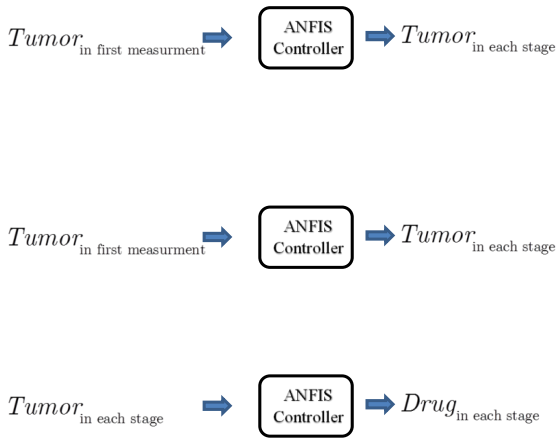

**Figure 2** ANFIS Control Diagram

**4.2.2 ANFIS observer**
In this work, an ANFIS observer is designed to estimate the other necessary states of the basic dynamic system. Since the behavior of immune cells, as we see in equation (1), depends on the tumor to train ANFIS observers to estimate safe cells, the tumor is considered to be an input variable and the output variables are immune cells. These results give specialists ability to predict the behavior of the immune system during treatment.

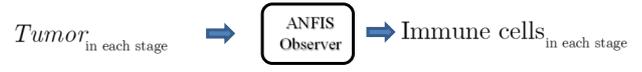

**Figure 3** ANFIS Observer to estimate safe cells

Obviously to observe the number of normal cells by considering equation (3) Tumor should be consider as an input variable.

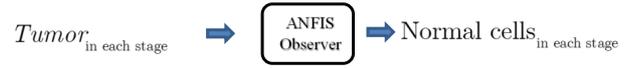

**Figure 4** ANFIS Observer to estimate infected cells

To train these ANFIS observers several different patients were considered.

**5.Results**
This section examined the capabilities of the ANFIS controller and the observer in two different patients and compared the discrete results of ANFIS controllers and observers with the optimal continuous controller. Initial conditions are shown in Table 2.

**Table 2** Initial conditions

| parameters | Value |
|---|---|
| I (0) | 0.15 |
| T (0) | 0.2 |
| N (0) | 1 |
| F (0) | 0.25 |
| M (0) | 0.25 |

**5.1 ANFIS Controller result**
This figure shows the ability of our ANFIS controller to estimate the amount or volume of Tumor just by measurement on the first day.

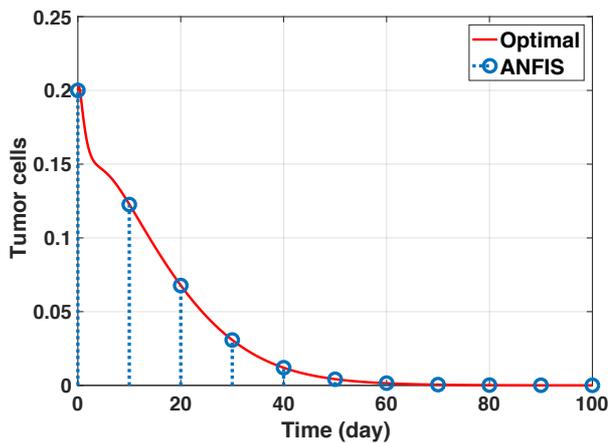

**Figure 5** Comparison the discrete amount or volume of Tumor by ANFIS controller with real data

The high ability of discrete ANFIS controller, designed and trained to predict the optimal amount of drug was shown by comparison with the result of optimal continuous controller.

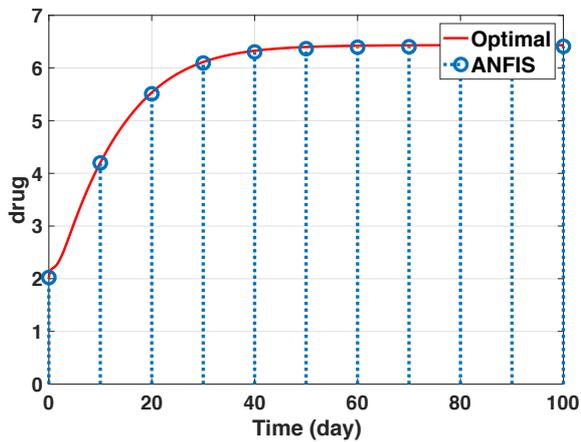

**Figure 6** Comparison the amount of discrete drug for patient by ANFIS controller with continuous amount of drug by optimal controller

The discrete results of the ANFIS controller are very similar to the continuous results of the optimal controller, although the ANFIS controller only needs to know the amount or volume of the tumor on the first day of treatment.

The next graph contained the amount of drug in these two controls, but this time for another patient.

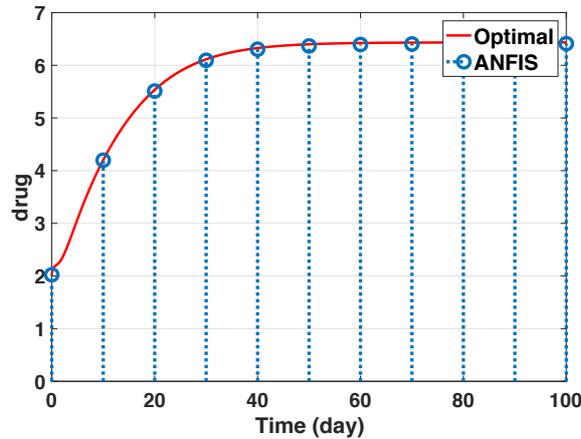

**Figure 7** Comparison the amount of discrete drug for patient by ANFIS controller with continuous amount of drug by optimal controller for new patient

This kind of discrete and roughly estimated amount of the drug can be really helpful for specialists to predict the situation they can meet for each patient under different conditions.

The key point is that the ANFIS controller is not a model base and can work with experimental data.

### 5.2 ANFIS observer results
#### 5.2.1 Monitoring immune cells

In this figure, the trend immune cells were compared with experimental data during treatment with ANFIS observer.

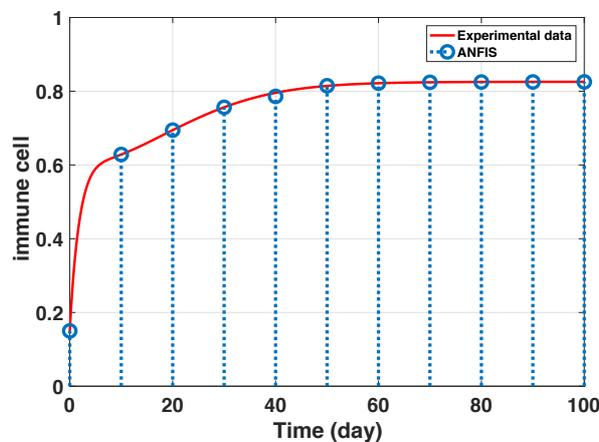

**Figure 8** Comparison the amount of discrete number of immune cells for patient by ANFIS controller with continuous amount of drug by optimal controller

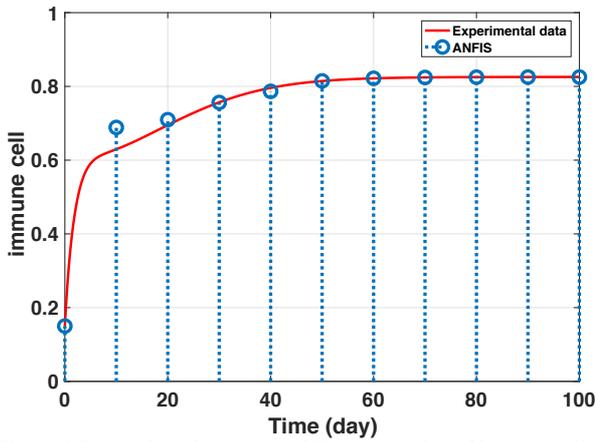

**Figure 9** Comparison the amount of discrete number of immune cells for patient by ANFIS controller with continuous amount of drug by optimal controller for new patient

Although error can be seen in the early days, results show that ANFIS observer can accurately estimate behavior of immune cells during treatment for new patients in the course of time.

### 5.2.2 Monitoring safe cells

In figure 10, the trend normal cells during the treatment with ANFIS observer was compared to experimental data.

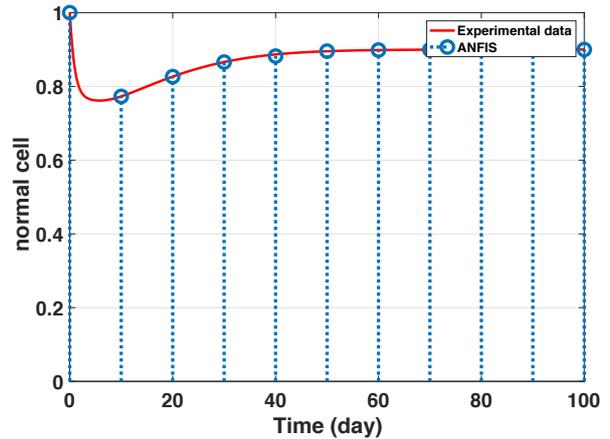

**Figure 10** Comparison the amount of discrete number of immune cells for patient by ANFIS controller with continuous amount of drug by optimal controller

The comparison was repeated for new patient.

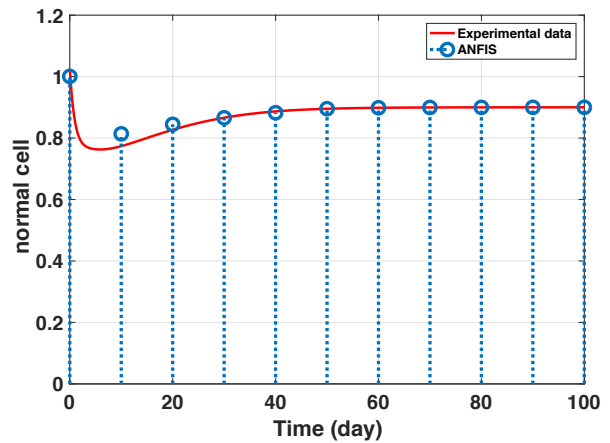

**Figure 11** Comparison the amount of discrete number of immune cells for patient by ANFIS controller with continuous amount of drug by optimal controller for new patient

The high ability of ANFIS observer leads to the monitoring of the other necessary state of control system in each treatment section.

### 6. Conclusion
In this study, an ANFIS (adaptive neuro-fuzzy inference system) controller for treating cancer was presented, and an ANFIS observer rendered to estimate the necessary system conditions. The ANFIS controller only needs the amount or volume of tumor cells as input variables to discretely estimate the amount of drug needed in order to eliminate tumor by minimizing the number of healthy cells which would be killed during the treatment. This controller can provide treatment schedule similar to what specialists do in the clinical environment, which of course is followed by a strong link between medical and engineering sciences. This goal was achieved by using real patient data to directly train ANFIS controllers. However, in previous models, all necessary states including healthy and immune cells should be known to estimate dose of the drug.

In simulations, the amount of tumor cells in a patient was measured in the first day of treatment, and then the ANFIS system separately reported information about the amount or volume of the tumor and the appropriate dose of the drug during the treatment period for each specific patient. The performance of the ANFIS controller was compared with the optimal PID controller and the performance of the ANFIS observer compared to the experimental data. The accuracy of our results shows that our proposed method can be used in realistic clinical treatments that help optimize the dose of the drug injected per day or week and consequently decrease the treatment period and healthy cells kill rate.

## References

[1]  R. L. Siegel and e. al, "Colorectal Cancer Statistics," *A Cancer Journal for Clinicians,* vol. 67, no. 3, pp. 177-193, 2017.

[2]  SM Abtahi and M Sharifi, "Machine Learning Method to Control and Observe for Treatment and Monitoring of Hepatitis B Virus," *arXiv:2004.09751,* 2020.

[3]  J. A. Florian, J. L. Eiseman and R. S. Parker, "A nonlinear model predictive control algorithm for breast cancer treatment," *IFAC Proceeding Volumes,* vol. 37, no. 9, pp. 929-940, 2004.

[4]  Hamed Moradi, Mojtaba Sharifi and Gholamreza Vossoughi, "Adaptive robust conrol of cancer cheomotherapy in the presence of parametric uncertainties: A comparison between three hypotheses," *Computers in Biology and Medicine,* vol. 56, pp. 145-157, 2015.

[5]  A. Yazdanpanah Goharrizi, R. H. Kwong and R. Chopra, "A self-tuning adaptive controller for 3-D image-guided ultrasound cancer therapy," *IEEE Transactions on Biomedical Engineering,* vol. 61, no. 3, pp. 911-919, 2014.

[6]  J. J. Cunningham, J. S. Brown, R. A. Gatenby and K. Staňková, "Optimal control to develop therapeutic strategies for metastatic castrate resistant prostate cancer," *Journal of Theoretical Biology,* vol. 459, pp. 67-78, 2018.

[7]  G. W. Swan, "Role of optimal control theory in cancer chemotherapy," *Mathematical Biosciences,* vol. 101, no. 2, pp. 237-284, 1990.

[8]  H. Moradi, G. R. Vossoughi and H. Salarieh, "Optimal robust control of drug delivery in cancer chemotherapy: a comparison between three control approaches," *Computer Methods and Programs Biomedicine,* vol. 112, p. 69–83, 2013.

[9]  Y. Batmani and H. Khaloozadeh, "Optimal chemotherapy in cancer treatment: state dependent Riccati equation control and extended Kalman filter," *Optimal Control Applications and Methods,* vol. 34, no. 5, pp. 562-577, 2013.

[10] Y. Batmani and H. Khaloozadeh, "Optimal chemotherapy in cancer treatment: state dependent Riccati equation control and extended Kalman filter," *Optimal Control Applications and Methods,* vol. 34, no. 5, pp. 562-577, 2012.

[11] E. Al-Daoud, "Cancer diagnosis using modified Fuzzy network," *Universal Journal of Computer Science and Engineering Technology,* vol. 1, no. 2, pp. 73-78, 2010.

[12] V. Balanica and e. al, "Evaluation of breast cancer risk by using Fuzzy logic," *U.P.B. Sci. Bull., Series C,* vol. 73, no. 1, 2011.

[13] R. El Hamidi, M. Njah and M. Chtourou, "An evolutionary neuro-fuzzy approach to breast cancer," in *Systems Man and Cybernetics (SMC), 2010 IEEE International Conference on*, Istanbul, 2010.

[14] J.-S. R. Jang, "ANFIS: adaptive-network-based fuzzy inference system," *ieee transaction on systems,* vol. 23, 1993.

[15] F. Schabel Jr., H. Skipper and W. Wilcox, "Experimental evaluation of potential anticancer agents. XIII. On the criteria and kinetics associated with curability of experimental leukemia," *Cancer Chemotherapy Report,* vol. 25, pp. 1-111, 1964.

[16] L. Norton and R. Simon, "Tumor size, sensitivity to therapy, and design of treatment schedules," *Cancer Treatment Reports,* vol. 61, pp. 1307-1317, 1977.

[17] A. d'Onofrio, A. Gandolfi and S. Gattoni, "The Norton–Simon hypothesis and the onset of non-genetic resistance to chemotherapy induced by stochastic fluctuations," *Journal of Physica A,* vol. 391, p. 6484–6496, 2012.

[18] N. Babaei and M. U. Salamci, "Personalized drug administration for cancer treatment using model refrence adaptive control," *Journal of Theoretical Biology,* vol. 371, pp. 24-44, 2015.

[19] R. Ku-Carrilo, S. E. Delgadillo and Chen-Charpentier, "A mathematical model for the effect of obesity on cancer growth and on the immune system response," *Applied Mathematical Modeling,* vol. 40, no. 7-8, pp. 4908-4920, 2016.

[20] Michio SUGENO and Tomohiro TAKAGI, "Fuzzy identification of systems and its applications to modeling and control," *fuzzy sets and systems,* vol. 15, pp. 116-132, 1985.